\documentclass[sigconf, nonacm]{acmart}

\usepackage{etoolbox,graphicx,setspace,listings,multicol,multirow,xspace,enumitem,booktabs,xcolor,subcaption}
\usepackage{listing}
\usepackage{amsmath}
\usepackage{algorithm}
\usepackage{algpseudocode}
\usepackage{comment}
\usepackage{balance}
\balance

\usepackage{tikz}
\usetikzlibrary{calc}

\usepackage{booktabs}
\usepackage{graphicx}
\usepackage{multirow}
\usepackage{makecell}

\newcommand*{\sn}{\textsc{Zelda}\xspace}
\newcommand*{\snverbose}{\textbf{Z}ero-shot
\textbf{E}xpressive, re\textbf{L}evant, and \textbf{D}iverse video \textbf{A}nalytics\xspace}
\newcommand*{\clip}{CLIP\xspace}

\newcommand*{\blrel}{\texttt{\clip-Relevant}\xspace}
\newcommand*{\bldiv}{\texttt{\clip-Diverse}\xspace}
\newcommand*{\vivaf}{\texttt{VIVA-Fast}\xspace}
\newcommand*{\vivaa}{\texttt{VIVA-Accurate}\xspace}
\newcommand*{\topk}{top-K\xspace}

\newcommand*{\perfaccurate}{$10.4\times$\xspace}
\newcommand*{\perffast}{$4.7\times$\xspace}
\newcommand*{\perfavg}{$7.5\times$\xspace}

\newcommand*{\apsoverbrel}{$1.16\times$\xspace}
\newcommand*{\apsoverbdiv}{$1.09\times$\xspace}

\newcommand*{\mapoverbdiv}{$1.15\times$\xspace}

\newcommand*{\dcleaners}{\q{blurry, grainy, low resolution, foggy, sepia}}

\newcommand*{\introq}{\q{cars during daytime at traffic intersections\xspace}}

\setlist[itemize]{leftmargin=*}
\setlist[enumerate]{leftmargin=*}

\newcommand*{\eg}{\emph{e.g.,}~}
\newcommand*{\ie}{\emph{i.e.,}~}

\newcommand*{\msec}[1]{Section~\ref{sec:#1}\@\xspace}

\newcommand*{\tworowsc}[2]{\begin{tabular}[c]{@{}c@{}}#1\\#2\end{tabular}}
\newcommand*{\threerowsc}[3]{\begin{tabular}[c]{@{}c@{}}#1\\#2\\#3\end{tabular}}
\newcommand*{\tworowslt}[2]{\renewcommand\arraystretch{0.8}\begin{tabular}[l]{@{}l@{}}\vspace{-6pt}\\#1\\\vspace{2pt}#2\end{tabular}}
\newcommand*{\onerowlt}[1]{\renewcommand\arraystretch{0.8}\begin{tabular}[l]{@{}l@{}}\vspace{-6pt}\\#1\vspace{2pt}\end{tabular}}

\newcommand{\minihead}[1]{{\vspace{.45em}\noindent\textbf{#1.~~}}}
\newcommand{\miniheadit}[1]{{\vspace{.45em}\noindent\textit{#1.~~} }}

\newcommand{\kt}[1]{{\vspace{.45em}\noindent{\textit{Key Takeaway~--\xspace#1}}}}

\newcommand{\q}[1]{``#1''}

\usepackage{etoolbox} %

\begin{document}

\title{\sn: Video Analytics using Vision-Language Models}

\author{Francisco Romero}
\authornote{Denotes equal contribution.}
\affiliation{%
    \institution{Stanford University}
    \country{}
}
\email{faromero@stanford.edu}
\author{Caleb Winston}
\authornotemark[1]
\affiliation{%
    \institution{Stanford University}
    \country{}
}
\email{calebwin@stanford.edu}
\author{Johann Hauswald}
\affiliation{%
  \institution{Stanford University}
    \country{}
}
\email{johannh@stanford.edu}
\author{Matei Zaharia}
\affiliation{%
    \institution{Stanford University}
    \country{}
}
\email{matei@cs.stanford.edu}
\author{Christos Kozyrakis}
\affiliation{%
    \institution{Stanford University}
    \country{}
}
\email{christos@cs.stanford.edu}

\begin{abstract}
  Advances in ML have motivated the design of video analytics systems
  that allow for structured queries over video datasets. 
  However, existing systems limit query expressivity, require
  users to specify an ML model per predicate, rely on complex optimizations that
  trade off accuracy for performance, and return large amounts of redundant and
  low-quality results. This paper focuses on the recently
  developed Vision-Language Models (VLMs) that allow users to query images using
  natural language like \q{cars during daytime at traffic intersections.}
  Through an in-depth analysis, we show VLMs address
  three limitations of current video analytics systems: general expressivity,
  a single general purpose model to query many predicates, and are both simple
  and fast.
  However, VLMs still return large numbers of redundant and low-quality results
  that can overwhelm and burden users.
  In addition, VLMs often require manual prompt engineering to improve result
  relevance.

  We present \sn: a video analytics system that uses VLMs to return
  both relevant and semantically diverse results for \topk queries on
  large video datasets.
  \sn prompts the VLM with the user's query in natural language.
  \sn then automatically adds discriminator and synonym terms to boost
  accuracy, and terms to identify low-quality frames.
  To improve result diversity, \sn uses semantic-rich VLM embeddings in an
  algorithm that prunes similar frames while considering their relevance to the
  query and the number of \topk results requested.
  We evaluate \sn across five datasets and 19 queries and quantitatively
  show it achieves higher mean average precision (up to \mapoverbdiv)
  and improves average pairwise similarity (up to \apsoverbrel) compared
  to using VLMs out-of-the-box. We also compare \sn to a
  state-of-the-art video analytics engine and show that \sn retrieves
  results \perfavg (up to \perfaccurate) faster for the same accuracy and
  frame diversity.
\end{abstract}

\maketitle

\begin{picture}(0,0)
  \put(30.9,489.1){\hbox{\includegraphics[scale=0.0025]{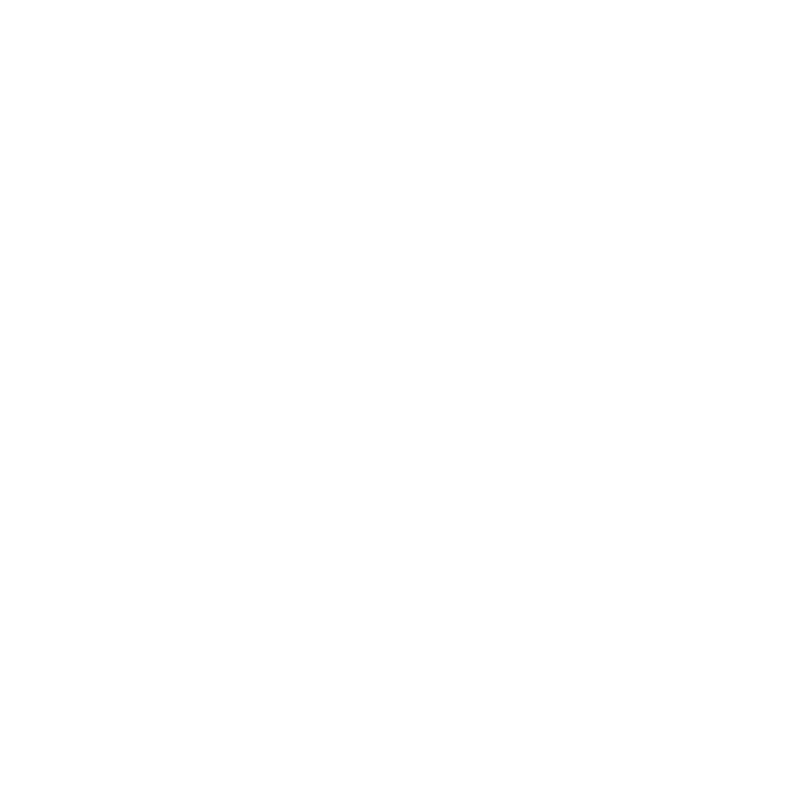}}}
\end{picture}

\vspace*{-\baselineskip}
\section{Introduction} \label{sec:intro}
Video analytics systems have become a topic of significant interest due to the
availability of vast video collections and the recent advances in ML.
Systems like VIVA~\cite{viva}, EVA~\cite{eva}, BlazeIt~\cite{blazeit}, and
others~\cite{figo, noscope, everest} allow users to query videos for objects,
individuals, actions, or complex scenes by specifying a list of predicates
that apply ML models on video frames.
However, as we show in Figure~\ref{fig:intro-motivation}a, today's
video analytics systems are limited in the following ways:
\begin{itemize}
  \item \textbf{Limited Expressivity}: Most existing video analytics systems
    limit query predicates to the trained classes of the available ML model(s).
\newcommand{\fig}{tex/figures/intro-diagram-daytime.pdf}

\begin{figure}[H]
  \centering
  \begin{minipage}[h]{1\columnwidth}
    \centering
    \includegraphics[width=1.0\columnwidth]{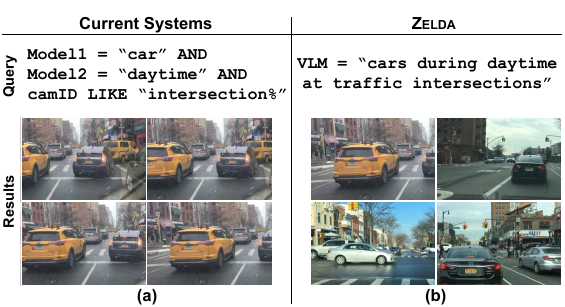}
    \caption{
      Current systems (a) limit predicate expressivity, require user expertise
      for model selection, and are complex to use.
      Our system \sn (b) allows expressive querying in natural
      language using a single fast model while returning relevant and diverse results.
    }
    \label{fig:intro-motivation}
  \end{minipage}
\end{figure}

    Some systems like VOCAL~\cite{vocal, vocalexplore} and SeeSaw~\cite{seesaw}
    require human-in-the-loop annotations to fine-tune models on new classes.
    A user is unable to query for \q{convertible} if the model is only trained on
    \q{cars} unless they augment the model with \q{convertible} training data or
    approximately map cars to convertibles, which is error-prone.
  \item \textbf{Specify Model per Predicate}: Existing video analytics systems
    require users to match query predicates to models.
    Users have to train a new model if none exist for their desired
    predicate(s).
    In Figure~\ref{fig:intro-motivation}a, a user must specify the models to use
    for finding \q{car} and \q{daytime} for the query \q{cars during daytime at
    traffic intersections.}
    Then, they need to reason about whether to use metadata like the camera
    location or train an additional model to filter by traffic intersections.
  \item \textbf{Complex Performance/Accuracy Tradeoffs}: Systems like
    BlazeIt, NoScope, VIVA, and FiGO trade off performance for accuracy by using
    model variants.
    Other systems like Everest~\cite{everest} and Probabilistic
    Predicates~\cite{pp} train query-specific proxy models on the critical path
    of a query.
    A system's query optimizer may choose to use a faster but less accurate car
    classifier, or train a query-specific binary classifier for daytime instead
    of using a larger model.
    These techniques make current systems significantly more complex.
  \item \textbf{Redundant and Low-Quality Results}: When querying large video datasets,
    a large number of frames may match the user's query.
    Many of these results will look visually similar and lack diversity.
    Existing systems will produce outputs similar to
    Figure~\ref{fig:intro-motivation}a, where the same car at different
    timesteps is ranked highest based on the model's confidence.
    These results may also contain low-quality blurry results. Needing to
    manually remove low-quality and redundant results unnecessarily burdens
    users.
  \end{itemize}
\noindent
Vision-Language Models (VLMs) are a new class of ML models for tasks such as
image classification (\clip~\cite{clip}), object detection (Detic~\cite{detic}
and ViLD~\cite{vild}), and video understanding (VideoCLIP~\cite{videoclip}).
VLMs jointly represent natural language text (known as prompts) and images
in the same embedding space.
The cosine distance of prompt and image embeddings
represents how semantically similar they are~\cite{hfblog,clipwebsite}.
The general architecture of VLMs is shown in Figure~\ref{fig:vlm-arch}.
These models are pretrained on large datasets of captioned images with the
goal of maximizing the similarity between text-image pairs that are
semantically related, while minimizing the similarity of non-corresponding
pairs~\cite{laion}. The resulting embeddings have been shown to encode rich and
complex semantic information about images~\cite{clip,hfblog,detic,vild}.
In this paper, we advocate for using VLMs to address the limitations of
existing video analytics systems.
Specifically, we perform an in-depth analysis across a wide range of datasets
and queries to demonstrate VLMs help to address the first three challenges:
\begin{itemize}
  \item \textbf{General Expressivity}:
    Users can express a wide range of predicates because VLMs understand
    natural language. For example, \q{car} and \q{convertible} are close in
    embedding space allowing a VLM to retrieve semantically similar frames.
  \item \textbf{A General Purpose Model}:
    VLMs are zero-shot: they do not have a pre-configured set of classes making
    them applicable for a wide range of queries.
    VLMs can be used out-of-the-box as a general purpose model with high
    accuracy or alongside special-purpose models (\eg find specific car types).
  \item \textbf{Simple and Fast}:
    Using a single type of model reduces system complexity and simplifies a
    data analyst's workflow.
    This eliminates the need to manage model variants or employ additional
    logic to train query-specific models.
    Some VLMs are fast (1000s of frames per second on an NVIDIA T4
    GPU~\cite{nvidiat4}) since they only require embedding comparisons.
\end{itemize}

\noindent
However, VLMs alone still face two challenges.
First, as is the case with most generative AI models, VLMs need careful
prompting to be sufficiently accurate.
For example, the query \q{car crash} can be expressed in many different ways:
\q{cars on fire}, \q{car pileup}, or \q{damaged car}.
There have been several proposed approaches for automating prompt engineering,
especially for large language models~\cite{autoprompt}.
These methods tune a single prompt for more accurate text generation.
However, classification for video analysis requires multiple prompts to
discriminate different classes.
Second, current systems overwhelm users with large numbers of frames that may be
too similar or low-quality.
While systems like NoScope use pixel-wise difference detectors to diversify
results, they require per-query tuning which is error-prone, and only use pixel
differences instead of richer semantic information.
Systems like VOCALExplore require users to manually identify low-quality
frames.
This is burdensome to users, especially since it must be repeated for every new
video dataset they wish to explore.
Everest limits results presented to users but does nothing to address diversity
or quality of their results.

We present \sn, a system for \snverbose. \sn uses VLMs to generate
relevant and semantically diverse results for \topk queries on large
video datasets.
As shown in Figure~\ref{fig:intro-motivation}b, \sn takes natural
language text as input and generates candidate frames without any user
annotations or model fine-tuning.
To increase accuracy, we develop the right VLM prompting strategy for video
analytics.
In addition to the user's query, \sn automatically prompts the VLM with three
sets of terms:
(1) synonyms to surface all possible results that can be relevant to a query
(\eg \q{car crash} will also include \q{car rollover} results),
(2) a diverse set of labels that act as discriminators, and
(3) terms that help identify low-quality frames like \q{blurry, grainy.}
\sn removes low-quality frames by determining when that VLM is more confident
that a term like \q{blurry} matches a frame compared to the user's query.
To score candidates for semantic similarity, \sn uses precomputed VLM
embeddings with the diversity term introduced by the Maximal Marginal Relevance
(MMR) algorithm~\cite{mmr}.
This allows \sn to prune frames that are too similar while considering their
relevance to the query and the number of results requested by the user.
Finally, \sn returns the \topk frames that are
both relevant and semantically diverse.

We built \sn on top VIVA~\cite{hints}, an open-source query engine for
video analytics, using \clip developed by OpenAI~\cite{clip}.
We evaluated \sn using five video datasets and 19 queries that feature
both single and multiple predicates. We show that \sn's ranking
pipeline has \apsoverbrel better average pairwise similarity (APS) and
up to \mapoverbdiv higher retrieval mean average precision (MAP)
compared to using a VLM out-of-the-box. We also compare \sn to
VIVA~\cite{hints}, a state-of-the-art baseline for \topk video
analytics, and show that \sn retrieves results \perfavg
(up to \perfaccurate) faster for the same accuracy and frame diversity.

In summary we make the following contributions:
\begin{itemize}
\item We highlight the potential of applying VLMs to video analytics
  showing their ease of use, high accuracy, and performance benefits
  across diverse set of
  queries~(Sections~\ref{sec:motivation-workflow}
  and~\ref{sec:motivation-ootb}).
\item We identify the limitations inhibiting VLMs from returning
  out-of-the-box relevant and diverse results~(\msec{topk}).
\item We present \sn, a video analytics system that uses VLMs to
  enable queries in natural language and automatically produce high quality,
  relevant, and semantically diverse results by applying the right prompting
  strategy~(\msec{system}).
\item We quantitatively show that \sn improves result diversity
  compared to just using VLMs out-of-the-box. We demonstrate
  that by leveraging VLMs, \sn simplifies query optimization and
  improves performance over existing state-of-the-art video analytics
  systems~(\msec{evaluation}).
\end{itemize}

\section{Vision-Language Model Primer} \label{sec:background}
\begin{figure}[t]
  \centering
  \begin{minipage}[h]{1\linewidth}
    \centering
    \includegraphics[width=1\linewidth]{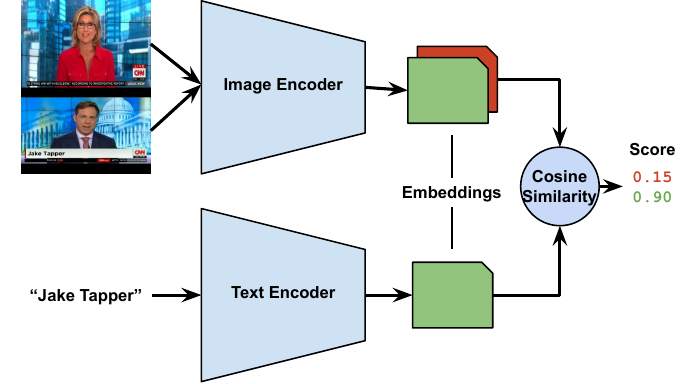}
    \caption{
      General architecture of vision-language models.
    }
    \label{fig:vlm-arch}
  \end{minipage}
\end{figure}

A Vision-Language Model (VLM)\footnote{VLMs are also referred to as Image
Language Models (ILMs). We use VLM throughout this work without loss of
generality.} is a type of ML model that is
trained on image-text pairs such as an image of Jake Tapper with the
caption \q{Jake Tapper, a news anchor}~\cite{hfblog,laionknn}.
The model learns from complex scenes and their natural language descriptions
to encode visual and semantic information.
As shown in Figure~\ref{fig:vlm-arch}, the model is composed of an image encoder
and a text encoder.
These encoders generate images and text vector embeddings, relying on
transformers to capture the sequence-to-sequence nature of both language and
vision~\cite{attention, vit}.
The goal of the model's image and text encoders is to learn
similar embeddings for similar image-text pairs.
The embeddings can be used to estimate similarity between image and
text data. The embeddings can also be used for text-text and image-image
similarity analysis.
VLMs are pretrained on very large datasets of image-text pairs mined from the
Internet, which allows them to be used zero-shot (\ie without
fine-tuning) on a variety of downstream tasks~\cite{laion}.

VLMs can estimate the similarity of video frames to a given natural language
prompt.
The image and text encoders generate embeddings representing the video frames
and the input prompt, respectively.
Prompts often extend a template such as \q{a photo of \{prompt\}}
to assimilate a caption~\cite{clipprompts}.
Multiple prompts may be compared to one or more frames.
The frame and prompt embeddings are compared using cosine similarity where
frame-prompt pairs with highest similarities are most likely to be related, as
shown in Figure~\ref{fig:vlm-arch}.
Computing frame and prompt embeddings, and their cosine similarities is a
fast operation. The cosine similarities use highly optimized vector search
libraries like FAISS developed by Meta~\cite{faiss} or vector search databases like
PineconeDB~\cite{pinecone}. In~\msec{motivation-perf}, we compare the
out-of-the-box performance of a VLM to models commonly used in video analytics.

\clip is an example of a VLM that achieves state-of-the-art accuracy
on zero-shot image classification~\cite{clip}. \clip was trained on 400
million image-text pairs mined from the Internet.
Other VLMs like ALIGN~\cite{align} and DeCLIP~\cite{declip} show similar
accuracies to \clip using noisy or less training data, respectively.
VLMs have also been used for object detection~\cite{detic, vild},
segmentation~\cite{segment},
visual-question answering and captioning~\cite{flamingo, blip2}, image-text
retrieval~\cite{filip}, and video understanding~\cite{videoclip,clip4clip}. Our work aims
to take advantage of the unique characteristics of VLMs as general image and
text encoders and will benefit from the evolving research in this field.

\section{VLMs for Video Analytics} \label{sec:motivation}
In this section, we explore applying VLMs to video analytics. We first show how
a user would query videos using a VLM-based system compared to current systems.
We then quantitatively and qualitatively evaluate the accuracy and throughput of
VLMs compared to ML models commonly existing in video analytics systems.
Finally, we identify the challenges limiting the use of VLMs for
\topk video analytics queries.
\subsection{Query Workflow} \label{sec:motivation-workflow}
We compare the video analytics query workflow using systems like
VIVA~\cite{viva} and EVA~\cite{eva} to a VLM-based system.
We consider Figure~\ref{fig:intro-motivation}'s query, where an 
analyst wants to find \introq.

\subsubsection*{Current Systems}
Since this query has three predicates --- \q{cars}, \q{daytime}, and
\q{traffic intersections} --- the analyst needs to use three models.
Commonly-used object detectors are trained to identify cars and are
typically used in existing video analytics systems.
However, finding models for \q{daytime} and \q{traffic intersections} is more
challenging. To detect daytime scenes, the analyst can either label frames
with day and night to train a small binary classifier or, if accurate metadata
is available, use timestamps to filter by daytime.
Similarly, the analyst could either train a model to identify intersections by
labeling training data or use camera metadata if they include location
information. In Figure~\ref{fig:intro-motivation}, we assume the analyst has a
model for daytime and metadata for intersections.  This workflow is burdensome
and error-prone for the analyst: it often requires manual labeling and model
training, or carefully validating available metadata. These issues must be
revisited every time the query changes.

\subsubsection*{VLM System}
The analyst simply prompts the VLM with the natural language query
\introq. The VLM has been trained
on many examples of cars, daytime scenes, and traffic intersections so
it can find likely matches without the need for a specialized model or
timestamps. This requires significantly less work from the analyst than
what current systems present since there is no need map predicates to
models or label data to train new models. The results using this
system are shown in Figure~\ref{fig:intro-motivation}b. We note the
results are visually different because we designed a VLM-based system
that removes low-quality frames and considers semantic diversity in
the results returned (further described in~\msec{system}).

\kt{
  Current systems rely on users' expert knowledge about what models
  to use for a given predicate, or when to use other sources of
  information like camera metadata. In contrast, a VLM-based system
  allows users to focus on expressing their query in natural language
  and leaves it to the system to return relevant results.
}

\subsection{Out-of-the-box Accuracy and Throughput} \label{sec:motivation-ootb}
We evaluate the out-of-the-box accuracy and performance of VLMs with \clip as
our reference VLM. While \clip has shown impressive zero-shot results on a range
of image classification tasks~\cite{clip}, we compare its accuracy against
predicates and models commonly used for video analytics. We consider two scenarios with single
and multiple predicates. In current video analytics systems, queries with
multiple predicates introduce additional complexity in selecting the best order
for model execution~\cite{hints,eva}. We show a single VLM prompt is just as
effective as a query with multiple predicates. We consider 3 different
tasks: traffic analysis on dashcam footage~\cite{bdd}, animal recognition on
trap camera footage~\cite{cct}, and action recognition on a popular action
detection dataset~\cite{ucf}.

\minihead{Single Predicate Accuracy}
For the three tasks we study, we select queries where there is an accurate
reference model and other queries where there is no such reference model
available (marked NR=No Reference). The tasks and predicates studied are
summarized in Table~\ref{tab:ootb-perf}.
An accurate reference model includes the query
predicate as one of the classes it was trained on. For traffic
analysis and animal recognition, we compare to a slow but accurate
\q{YOLO-accurate} objection detection model~\cite{yolov5} commonly
used in video analytics~\cite{noscope,blazeit,figo,boggart,everest}.
For the action recognition, we compare to SlowFast, a state-of-the-art
model recently published and used in prior work on video
analytics~\cite{torchaction, hints}. For each task, the VLM is prompted
with all the classes of the reference model. For example, YOLO is
trained on coco80~\cite{coco80}, a set of 80 common classes (\eg
\q{car, truck, person, dog, cat}). We evaluate top-5 F1 accuracy\footnotemark
computed using ground truth labels and averaged across all queries. For action
recognition, we use top-1 F1 because SlowFast only emits a single label per
group of frames. Using the top-1 or top-5 F1 score is consistent with prior
research in VLMs~\cite{clip}.

Our results in Table~\ref{tab:ootb-perf} show that \clip has
competitive accuracy with the SlowFast and YOLO models.
For two of three tasks where a reference model is available, \clip
outperforms the competitive baselines. YOLO-accurate has better
accuracy for traffic analysis, which is expected as it was trained for
classes common in traffic videos.
When no reference model (\q{NR}) is available (\eg for predicates like
\q{motorcycle rider}), \clip is still able to identify some results. We
highlight this scenario for cases where an analyst does not have a model
available and can still use a VLM to identify some results. These results can
either suffice for the analyst or can be used as a starting point to train a
model for less common classes.

\minihead{Multiple Predicate Accuracy} \label{sec:motivation-prompt}
Next, we evaluate the accuracy of VLMs on queries containing multiple predicates
like \q{pedestrians next to car at traffic light}.
Current systems require either the user or a system like Galois~\cite{galois} to
split the query into multiple predicates where model is executed for each
predicate and the merged results are returned to the user.
In a VLM-based system, a VLM can be prompted with either multiple predicates
like \q{pedestrian $\cap$ car $\cap$ traffic light} where the result is the
intersection (or the union) of frames that contain each predicate.
We take the union or intersection of frames based on the original intent of the
query. We compare this to passing the entire query \q{pedestrians next to car at
traffic light} as a single predicate to the VLM.

Our results in Table~\ref{tab:ootb-predicates} compare the top-5 F1 VLM accuracy
with single and multiple predicates. For queries with multiple predicates (\eg
\q{car at night at traffic intersections}), parsing the query into separate
predicates (\eg \q{cars $\cap$ night $\cap$ traffic intersections}) does not
improve accuracy. For queries like \q{vehicle}, parsing into multiple
predicates for different subordinate terms \q{car $\cup$ truck $\cup$ bus
$\cup~\dots$} slightly improves accuracy. We generally found a single predicate
achieves high accuracy. This further highlights the expressivity VLMs provide
since users can submit an entire query without having to reason about breaking
down the query into multiple predicates.

\begin{figure*}[t]
  \centering
  \begin{minipage}[t]{1\linewidth}
    \centering
    \includegraphics[width=0.8\columnwidth]{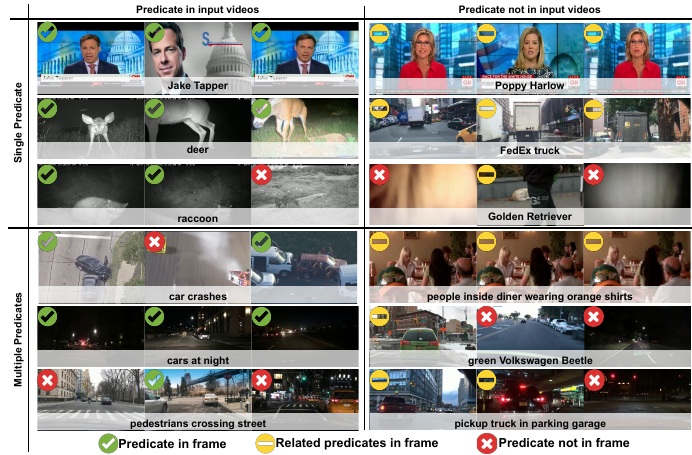}
    \caption{
      Top-3 VLM results for various queries.
      We study two cases: predicates in input videos and predicates not in
      input videos.
      To differentiate between \q{Predicate not in frame} and \q{Related
      predicates in frame}, we make a subjective assessment.
      For example, seeing trucks is useful for \q{FedEx truck}, while
      blurry frames for \q{Golden Retriever} is not.
      In both cases studied, VLMs return relevant results.
    }
    \label{fig:ootb-qualitative}
  \end{minipage}
\end{figure*}

\minihead{Qualitative Results} Figure~\ref{fig:ootb-qualitative} shows the three
highest-scoring results for a variety of single and multiple predicate queries.
We consider two cases: the predicate is present in the input videos and the
predicate is not present in the input videos.
These results demonstrate \clip can distinguish animals, people, and actions
like crossing a street. Even for predicates not present in the input
video, \clip still finds close matches using the semantic information
encoded in the embeddings. For single predicates like \q{Poppy
Harlow}, \q{FedEx truck}, and \q{Golden Retriever}, \clip finds people,
delivery trucks, and dogs respectively. For multiple predicate queries like
\q{pickup truck in parking garage}, \clip finds pickup trucks and
parking garages. A VLM's ability to find close matches even when the
predicate is not present demonstrates the potential of its zero-shot
capability for video analytics.

\minihead{Performance} \label{sec:motivation-perf}
Table~\ref{tab:motivation-throughput} compares \clip's throughput to models
commonly used in video analytics.
We take these measurements on an NVIDIA T4 GPU~\cite{nvidiat4} and measure
throughput in frames per second (FPS).
All models are preloaded on the GPU and benefit from batching frames to the GPU.
For the VLM, the performance includes encoding a single prompt and 10K frames.
The cosine similarity is calculated using FAISS, a fast vector search library
developed by Meta~\cite{faiss}.
We compare to a fast but less accurate \q{YOLO-fast}, and the slow but more accurate
\q{YOLO-accurate} object detection model for the traffic analysis
and animal recognition tasks.
Both models are trained on the same classes and are used for both tasks.
For the action recognition task, we compare to SlowFast.
\clip's throughout is an order of magnitude better than both the SlowFast and
YOLO models.
We note the YOLO models provide bounding boxes of objects detected while \clip
assigns a single label per frame.
Prior work has also investigated making \clip an object detector by executing
it multiple times over subframes~\cite{detic,vild}.
\footnotetext{For F1 accuracy, top-K refers to the order of the label confidences,
not the order of the frames returned.}

\begin{table}[t]
  \centering
  \caption{
    Accuracy (F1) of VLMs compared to Reference models (Ref.) commonly used in video
    analytics.
    For traffic analysis and animal recognition, we use a YOLO object detector.
    For action recognition, we use SlowFast.
    NR=No Reference.
  }
  \resizebox{\linewidth}{!}{%
    \begin{tabular}{llcc}
      \toprule
      \textbf{Task (\# Predicates)}   & \textbf{Predicates} & \textbf{VLM} & \textbf{Ref.} \\
      \midrule
      Traffic analysis (8)             & cars, pedestrian, $\dots$            & 0.45            & \textbf{0.70}                 \\
      Traffic analysis (NR) (3)        & traffic sign, motorcycle rider      & \textbf{0.21}            & --                   \\
      Animal recognition (2)           & dog, cat                            & \textbf{0.21}            & 0.17                 \\
      Animal recognition (NR) (6)      & deer, raccoon, $\dots$              & \textbf{0.29}            & --                   \\
      Action recognition (25)       & baseball pitching, archery, $\dots$ & \textbf{0.65}            & 0.50                  \\
      \bottomrule
    \end{tabular}
  }
  \label{tab:ootb-perf}
\end{table}

\begin{table}[t]
  \centering
  \caption{
    Accuracy (F1) of VLMs when running multiple predicates separately versus as
    a single predicate.
    VLMs achieve similar accuracies without needing to parse predicates.
  }
  \resizebox{1\linewidth}{!}{%
    \begin{tabular}{lc|lc}
      \toprule
      \textbf{Multiple Predicates} & \textbf{F1} & \textbf{Single Predicate} & \textbf{F1} \\ \hline
      \tworowslt{cars $\cap$ night}{$\cap$ traffic intersections} & 0.50 & \tworowslt{cars at night at}{traffic intersections} & \textbf{0.52} \\
      pedestrians $\cap$ street & 0.21 & \tworowslt{pedestrians}{crossing street} & \textbf{0.22} \\
      \onerowlt{car $\cup$ truck $\cup$ bus $\cup~\dots$} & \textbf{1.00} & \onerowlt{vehicle} & 0.88 \\
      \onerowlt{basketball $\cup$ fencing  $\cup~\dots$} & 0.72 & \onerowlt{sports} & \textbf{0.77} \\
      \bottomrule
    \end{tabular}
  }
  \label{tab:ootb-predicates}
\end{table}

\subsection{Analyzing Top-K Results} \label{sec:topk}
Section~\ref{sec:motivation-ootb} showed VLMs out-of-the-box are both accurate
and fast.
However, VLMs can overwhelm an analyst with results that are either redundant
or low-quality.
The result quality can also depend on the analyst's prompt.
Ideally, the analyst would like to review the top most relevant (answers the
query) and semantically diverse (variety in the predicate's context)
results.
Hence, we focus on improving the quality of \topk results in a VLM-based video
analytics systems.
Selecting the top-K results is a popular search functionality across many forms
of data~\cite{everest,topkdb,topksurvey,topx}.

Figures~\ref{fig:intro-motivation}a and \ref{fig:ootb-qualitative}
show what an analyst can expect using current systems or with VLM
out-of-the-box using cosine similarity ranking.
We highlight obvious limitations in the top-3 results of
Figure~\ref{fig:ootb-qualitative}.

\minihead{Low-Quality} Low-quality (\eg blurry) frames can be present
in a video dataset after decoding. For the query \q{Golden Retriever}, the first
and third results are blurry and indistinct. The second result shows a dog but
is partially occluded and distant. These frames rank highly when using a VLM's
cosine similarity metric but should not be prioritized over results where the
predicate is more visible.

\minihead{Duplicate Frames} Videos have high temporal similarity, even
when processed at a low frame rate (\eg 1FPS). Thus, while having
near-duplicate results are not surprising, they are not ideal when
processing large video datasets. Identical frames can also be
temporally distant. For example, a TV channel stream may include a
video sequence that is repeated when the same anchor is
introduced. For the query \q{Jake Tapper}, the first and third results are
nearly identical visually. The duplicate result should be ranked lower and
another frame should be prioritized where the predicate is in a different
setting (\eg Jake Tapper at a round table).

\minihead{Lack of Semantic Diversity} Showing semantically diverse
frames is important for interactive queries, where an analyst may
gradually refine the query as they see the results~\cite{eva}.
For the query \q{cars at night} in Figure~\ref{fig:ootb-qualitative}, all
results are visually distinct but semantically similar since they contain street
scenes. While all results are relevant, the lack of semantic diversity requires
the analyst to scan further to gather the same information. For example, with
greater semantic diversity, the top results for this query could show cars at
night on streets, highways, or bridges.

\minihead{Sensitivity to Prompt}
VLM results can depend on the quality of the input prompt.
For example, the query \q{car crash} only produced two relevant results.
However, we found that a similar prompt, \q{car rollover} would have produced
additional relevant results for an analyst to review.
Prompt engineering has become a prevalent challenge for models that take
natural language as an input (\eg large language models)~\cite{autoprompt}.
It requires analysts to spend additional time figuring out the right input(s)
to get the most relevant results.

\kt{
  VLMs enable expressive querying using natural language using a
  single model that is fast and accurate. This drastically reduces
  model management overheads and query development complexity.
  However, VLMs do not consistently provide relevant and semantically
  diverse \topk results out-of-the-box.
}

\begin{table}[t]
  \caption{
    Throughput of \clip on an NVIDIA T4 compared to commonly used video
    analytics models (measured with 10K 360p frames).
    \clip achieves up to 95.6$\times$ higher throughput.
  }
  \resizebox{0.8\linewidth}{!}{%
  \begin{tabular}{clc}
    \toprule
    \textbf{Tasks}          & \textbf{Model} & \textbf{FPS} \\
    \midrule
    \multirowcell{2}{Traffic analysis \\ Animal recognition}
                            & YOLO-fast     & \phantom{0}131     \\
                            & YOLO-accurate & \phantom{00}22     \\
    \midrule
    {Action recognition} & SlowFast      & \phantom{00}12     \\
    \midrule
    All                     & \clip         & 1149               \\
    \bottomrule
  \end{tabular}
}
  \label{tab:motivation-throughput}
\end{table}

\begin{figure*}[t]
  \centering
  \begin{minipage}[t]{1\linewidth}
    \centering
    \includegraphics[width=0.80\linewidth]{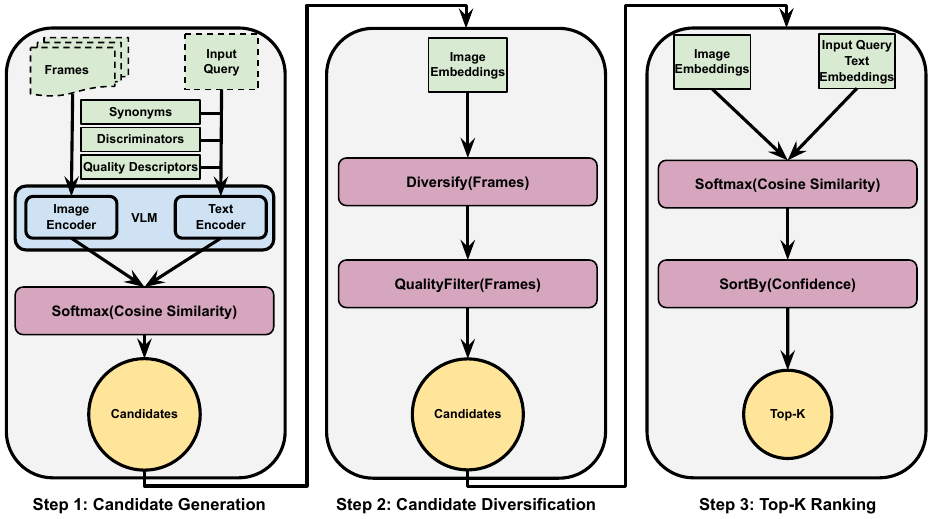}
    \caption{
      \sn's architecture: inputs are green (light), operations are 
      purple (dark), and results are yellow (circle).
      Dashed boxes are user-provided.
      (1) \sn generates candidate frames using VLM query scores, and
      discriminator and synonym terms,
      (2) \sn improves candidate diversity by removing similar frames and
      removing low-quality frames,
      (3) \sn ranks the final candidates to produce the \topk results.
    }
    \label{fig:arch}
  \end{minipage}
\end{figure*}

\section{\sn} \label{sec:system}
\sn is a system for \snverbose. It is designed to meet the following
requirements: 
\begin{enumerate}
  \item Retrieve relevant results given a natural language query input by
  applying the right prompting strategy.
  \item Prune low-quality and near duplicate frames, retaining a diverse
  set of candidate results.
  \item Return high confidence results that are semantically diverse.
\end{enumerate}
Existing \topk video analytics systems like Everest~\cite{everest} provide
probabilistic guarantees for \topk results but do nothing to improve diversity.
They are also limited to the classes provided by the model used (typically an
object detector).
Other video analytics systems like VIVA~\cite{viva}, EVA~\cite{eva}, and
BlazeIt~\cite{blazeit} do not offer support for \topk or diversifying results.
None of these systems provide a natural language query input interface.

Figure~\ref{fig:arch} illustrates \sn's architecture. %
\sn first uses the VLM to generate a candidate set of frames
(\msec{sys-cand-gen}).  Next, \sn improves result diversity and
quality using two techniques: (a) filtering to remove candidates that
are visually and semantically too similar to each other, and (b)
filtering to remove low-quality frames (\msec{sys-div}).
Finally, \sn ranks the remaining semantically diverse candidates
using the confidence of the VLM to generate the \topk results returned
to the user (\msec{sys-rank}). We now describe these components in detail.

\subsection{Prompting Strategy for Candidates} \label{sec:sys-cand-gen}
\sn's first step is to generate candidate frames to be considered for the final
\topk results. The user-provided inputs to candidate generation are video
frames and a natural language query.

As shown in Section~\ref{sec:motivation-ootb}, VLMs can provide accurate
results when prompted with just the classes that the user may query and using
the cosine distance to select relevant results.
However, we have observed two cases where just using the classes the user may
query is insufficient to produce accurate results.
First, there are queries where knowing how a VLM scores prompts relative to
each other can improve accuracy.
For example, for the query \q{cat}, other animals like a raccoon may look
similar.
Thus, knowing their relative scores to each other can help to better
\emph{discriminate} between the two.
Second, complex queries like \q{car crash} may only produce a subset of
relevant results.
By including the results of \emph{similar} terms like \q{car rollover}, more
candidate frames can be produced.

\sn applies an automated prompt strategy to improve the quality of candidates
generated.
To address the challenge of discriminating between classes that look similar,
we pass a large set of labels to \sn that serve as discriminators.
We use the 1,203 object categories from LVIS~\cite{lvis}.
LVIS is a recently-proposed dataset for studying instance segmentation
algorithms and has been used in recent VLM work for long-tail object
recognition~\cite{detic,vild}.
We found LVIS categories to be effective additional prompts across the wide
range of queries and datasets we studied because they represent every day
concepts that were carefully chosen to be distinct. \sn is not limited to LVIS
categories and alternative distinct label sets can be used.
To address the challenge of improving the relevant candidates, we pass 
synonym (\ie nearby) labels to \sn.
We generate the synonym terms using a language model (LM).
The LM is prompted with a request to generate $J$ synonyms of the input query
and \sn concatenates top-$K/J$ results for each synonym.
\sn uses Llama 2 70B~\cite{llama2} as the LM but is not limited to a particular
model.
We also pass \sn terms used to identify low-quality data using prompts
like \dcleaners.

\sn prompts the VLM's text encoder with (a) the user's query, (b) the distinct
label set classes, (c) synonym classes, and (d) the low-quality descriptors.
The input frames are fed into the VLM's image encoder.
Next, \sn computes the cosine similarity between the frame embeddings and the
embeddings generated by the text encoder.
For example, using LVIS as the distinct label set, two synonym classes, and five
low-quality descriptors, \sn will compute 1,211 cosine similarities per-frame.
The highest cosine similarity score from the synonym terms and the user's query
is then softmaxed with the rest of the cosine similarities to produce
confidence scores.
This enables \sn to compare the scores relative to each other.
\sn then ranks the frames by the softmax confidence score of the query
before passing them to the Candidate Diversification stage.
Similar to other ranking systems using language models like PLAID~\cite{plaid},
\sn does not impose a limit on the number of candidates since they will be
pruned in subsequent stages.
This also avoids the need for \sn to set or tune a threshold for how many
candidates to generate, which can be challenging when also considering frame
similarity and quality.

\subsection{Improving Result Diversity} \label{sec:sys-div}
Given the candidates produced by the Candidate Generation stage, \sn next uses
two techniques to improve result diversity while maintaining high 
accuracy: diversity filtering and quality filtering.
Diversifying video results for ease-of-annotation is important for training
vision models~\cite{encord2023}.

\minihead{Improving result diversity}
Videos have high temporal similarity where neighboring frames may only differ by
a small number of pixels. There can also be repeated segments that look
visually similar to each other like a TV news opening segment used multiple
times. Since VLMs (and ML models generally) produce similar confidences for
frames that are visually similar, there can be several highly-ranked but
redundant frames after the Candidate Generation stage. While this leads to high
accuracy \topk results, it is not ideal for users looking for a diverse set of
query results that can be further refined with their feedback.

\sn improves result diversity using Algorithm~\ref{alg:diversity}.
\sn visits the frames produced by the Candidate Generation stage in descending
ranked order (Line~\ref{alg:visit}).
For each frame, we compute pairwise cosine similarities with each of the other
previously-scored frames in \texttt{ScoredFrames}.
\sn assigns each frame a semantic similarity score that is its maximum pairwise
cosine similarity (Lines~\ref{alg:start-mpcs}-\ref{alg:end-mpcs}).
This step of the algorithm uses the second
term (diversity scoring) of the Maximum Marginal Relevance (MMR)
algorithm~\cite{mmr}. We drop the first term of MMR (relevance
scoring) because we already have similarity scores generated by the
VLM. MMR has been used extensively in information retrieval settings,
most notably in text summarization~\cite{mmrtext}.
The frames are then ranked based on their semantic similarity score
and pruned based on a system-set threshold that is query agnostic
(Line~\ref{alg:prune}). We empirically found a threshold of 0.80 preserves
relevant results while filtering those that are too similar to the top
confidence ones. We use this threshold for results presented in
Section~\ref{sec:evaluation}. To avoid over-filtering frames, \sn ensures there
are at least K frames available, regardless of the diversity score.

Existing video analytics systems, like NoScope~\cite{noscope} and
Boggart~\cite{boggart}, use visual similarity to reason about result
diversity. These systems build visual difference detectors
(VDDs), whose threshold must be set \emph{per-query}. Using visual
difference only removes frames whose pixels are nearby to each
other. Visual difference is not sufficient to reason about the
concepts in frames with respect to user's query (\eg cars versus
trucks). \sn is able to reason about semantic similarity because VLM
embeddings are trained such that more semantically similar frames have a higher
cosine similarity to each other. We demonstrate that using semantic similarity
improves the diversity of results in Section~\ref{sec:eval-vissim}.
\sn can be extended to use additional diversity functions. For example, using
the differences in labels assigned between two frames or systems like Everest
use object count to rank.

\begin{algorithm}[t]
  \small
  \caption{Result Diversity Algorithm}\label{alg:diversity}
  \begin{algorithmic}[1]
    \Procedure{DiversifyFrames}{$CandidateFrames, PruneThresh$}
    \State $ScoredFrames \gets []$
    \For{$\texttt{candidate in CandidateFrames}$}\label{alg:visit}
      \State $MaxCosSim \gets 0$
      \For{$\texttt{scored in ScoredFrames}$}\label{alg:start-mpcs}
        \State $CosSim \gets CosineSimilarity(candidate, scored)$
        \If{$CosSim > MaxCosSim$}
          \State $MaxCosSim \gets CosSim$
        \EndIf
      \EndFor
      \State $ScoredFrames.append((candidate, MaxCosSim))$\label{alg:end-mpcs}
    \EndFor
    \State \textbf{return} $PruneDiverse(ScoredFrames, PruneThresh)$\label{alg:prune}
\EndProcedure
\end{algorithmic}
\end{algorithm}

\minihead{Pruning low-quality frames}
Video often contains low-quality data like blurry or grainy frames.
Systems like VOCAL and VOCALExplore~\cite{vocal,vocalexplore} ask users to
manually identify low-quality data so it can be skipped when producing results.
Not only is this burdensome on users, but can diminish the quality of the \topk
results even if the frame contains the query.
As described in Section~\ref{sec:sys-cand-gen}, \sn automates low-quality data
removal by additionally prompting the VLM with terms such as \dcleaners.
Each of these terms is separately passed to the VLM's text encoder.
At the end of the Candidate Generation stage, \sn has produced the probability
confidence of these terms relative to the query and the label set by using a
softmax layer.
During the quality pruning stage, \sn uses this probability confidence to prune
frames that have a higher confidence than the query.
Like similarity pruning, \sn ensures there are at least K frames available
to be ranked.

\subsection{Selecting Top-K Results} \label{sec:sys-rank}
After pruning frames to improve result diversity in the Candidate Diversity
stage, \sn's final step is to produce the \topk ranked frames. To do so, \sn
ranks the remaining frames by the probability confidence of the input query and
returns K results.
This ensures of the remaining frames, the highest confidence
ones scored by the VLM will be shown first. By ranking after
diversifying frames, \sn produces highly accurate, relevant, and diverse results
that meet the following requirements:
\begin{itemize}
\item A diverse set of frames matching the query in the case when
  multiple high confidence frames are found. For example, when querying for a
  car in dashcam footage, \sn will produce different types and views of cars.
\item A diverse set of frames that are semantically similar when few high
  confidence frames are found. For example, when querying for baseball pitching
  in an action dataset, \sn will show matching frames while also including
  frames from related sports like cricket.
\end{itemize}

\subsection{\sn Implementation}
We implemented \sn using \clip as the VLM on top of the VIVA video
analytics engine~\cite{hints}.
VIVA is an open-source system for large-scale video analytics based on
Spark~\cite{spark}. VIVA optimizes complex query plans to meet user
accuracy goals by automatically applying user-specified, domain
knowledge about models. In implementing \sn, we leveraged VIVA's
common model registration interface to add \clip as the
reference VLM. This allowed us to define a user-defined function (UDF) that is
executed on Spark's workers.
We use VIVA and Spark's query optimizers out-of-the-box, which take
advantage of structured query optimizations for reading data and
assigning tasks to workers. After executing the VLM on the input frames, VIVA
invokes the \sn pipeline to select candidates, identify low-quality data and
similar frames, and return the \topk results.

The authors of \clip noted several limitations we also observed in
using \sn.
This includes lower accuracy for queries that
require fine-grained spatial understanding like \q{VW Golf turning
left at night}, counting objects in frames like \q{frames with 3
cars}, or very specific predicates like car brands~\cite{clip}.
VLMs are a fast advancing field and we expect their accuracy for such
queries will continue improving~\cite{align,declip,glip,unifiedio}.
Building \sn on VIVA allows us to combine both capabilities: using specialized
models alongside VLMs.
For example, after identifying cars in a complex scene described in natural
language, an object detector can be used to count cars, or a specialized model
can be used to identify cars with a specific brand's logo.
We leave this for future work.

\begin{table}[t]
  \centering
  \caption{
    Datasets, frames, and predicates.
    We use a mix of single and multiple predicate queries from recent work in
    video analytics.
  }
  \resizebox{\linewidth}{!}{%
    \begin{tabular}{llrc}
      \toprule
      \textbf{Dataset}                & \textbf{Description} & \textbf{Frames} & \textbf{Predicates} \\ \midrule
      BDD~\cite{bdd}                  & Dashcam footage  & 1000  & \threerowsc{vehicle, car,}{cars at night at traffic intersections,}{pedestrians crossing street}\\ \midrule
      CCT~\cite{cct}                  & Camera traps & 5000  & deer, raccoon \\ \midrule
      Movies~\cite{moviesdataset}     & Movie footage & 10000 & \tworowsc{vehicle, people inside diner,}{car, pedestrians crossing street}\\ \midrule
      News~\cite{tvarch21,cabletvnews}& News interviews & 10000 & \tworowsc{vehicle, Bernie Sanders,}{car, Jake Tapper}\\ \midrule
      UCF101~\cite{ucf}               & Action recognition & 4337  & \tworowsc{sports, baby, cake,}{baseball pitching, archery}\\
      \bottomrule
    \end{tabular}
  } %
  \label{tab:datasets}
\end{table}

\newcommand{\eesizef}{0.205}
\newcommand{\eesize}{0.190}
\newcommand{\mapfigdir}{tex/figures/e2e}

\begin{figure*}
  \centering
  \begin{subfigure}[t]{0.30\linewidth}
    \includegraphics[width=1\textwidth]{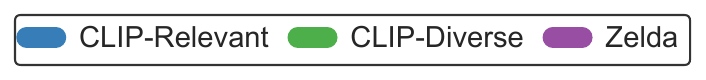}
  \end{subfigure}
  \\
  \begin{subfigure}[t]{\eesizef\linewidth}
    \includegraphics[width=1\textwidth]{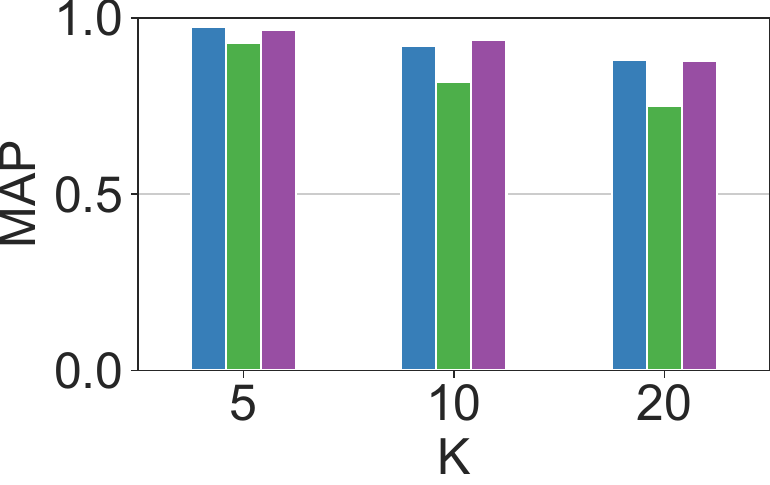}
    \caption{BDD}
    \label{fig:map-bdd}
  \end{subfigure}
  \begin{subfigure}[t]{\eesize\linewidth}
    \includegraphics[width=1\textwidth]{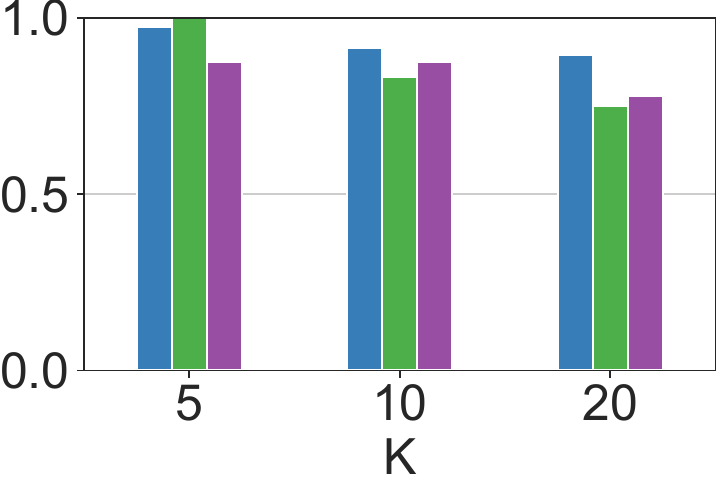}
    \caption{CCT}
    \label{fig:map-cct}
  \end{subfigure}
  \begin{subfigure}[t]{\eesize\linewidth}
    \includegraphics[width=1\textwidth]{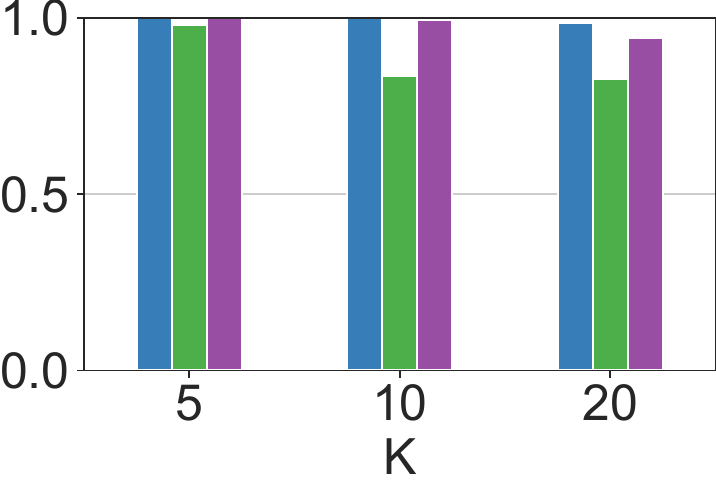}
    \caption{News}
    \label{fig:map-news}
  \end{subfigure}
  \begin{subfigure}[t]{\eesize\linewidth}
    \includegraphics[width=1\textwidth]{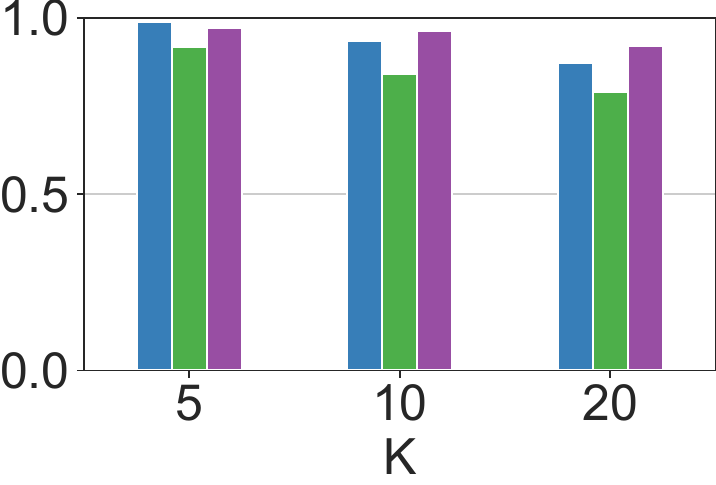}
    \caption{Movies}
    \label{fig:map-movies}
  \end{subfigure}
  \begin{subfigure}[t]{\eesize\linewidth}
    \includegraphics[width=1\textwidth]{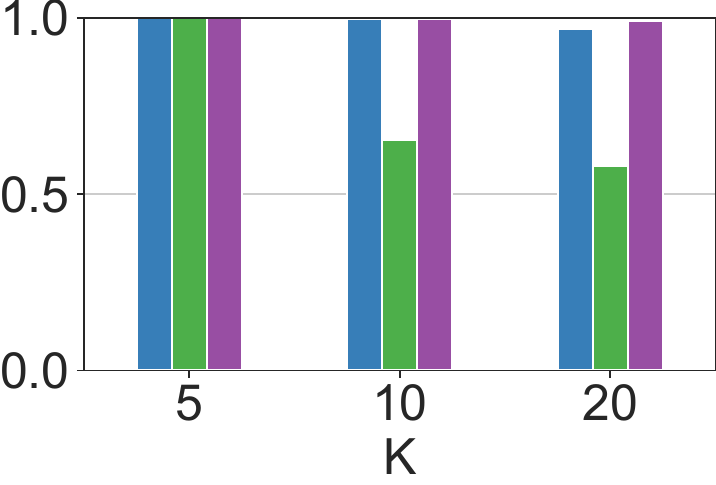}
    \caption{UCF101}
    \label{fig:map-ucf101}
  \end{subfigure}
  \caption{
    Mean average precision (MAP) for \sn and the baselines.
    Higher is better.
  }
  \label{fig:map-all}
\end{figure*}

\newcommand{\apsfigdir}{tex/figures/e2e}

\begin{figure*}
  \centering
  \begin{subfigure}[t]{\eesizef\linewidth}
    \includegraphics[width=1\textwidth]{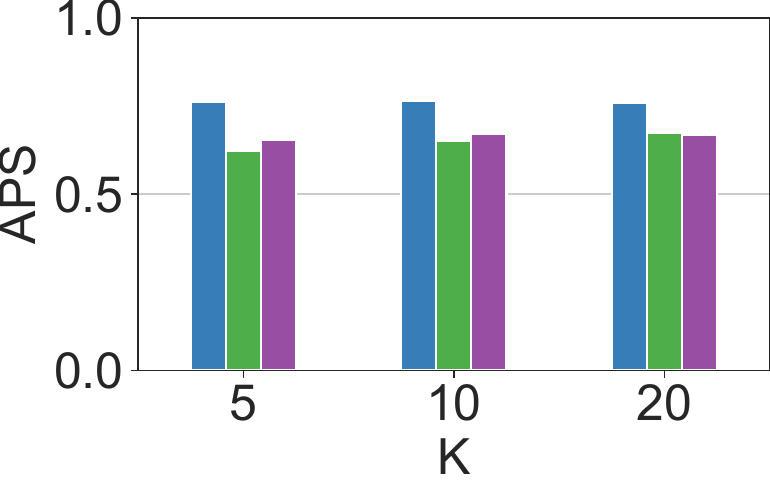}
    \caption{BDD}
    \label{fig:aps-bdd}
  \end{subfigure}
  \begin{subfigure}[t]{\eesize\linewidth}
    \includegraphics[width=1\textwidth]{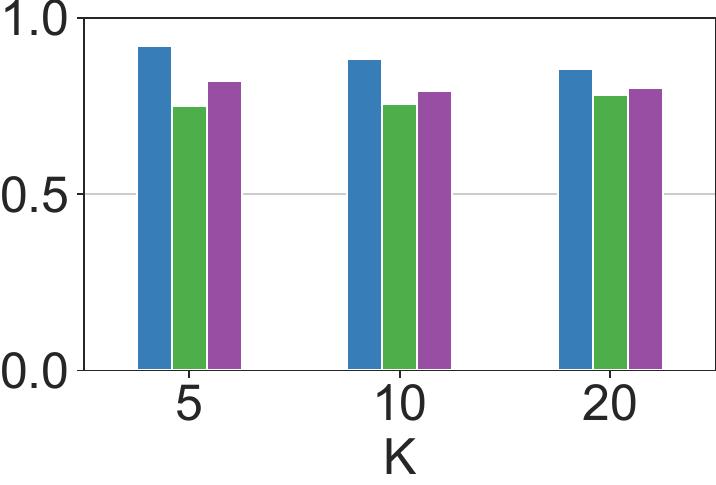}
    \caption{CCT}
    \label{fig:aps-cct}
  \end{subfigure}
  \begin{subfigure}[t]{\eesize\linewidth}
    \includegraphics[width=1\textwidth]{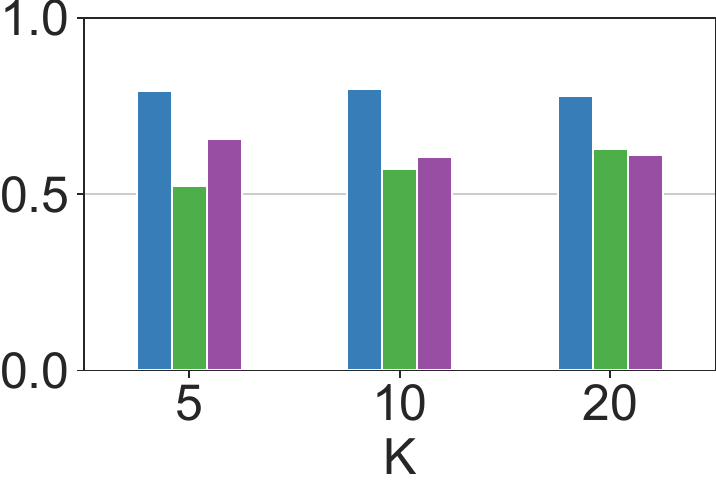}
    \caption{News}
    \label{fig:aps-news}
  \end{subfigure}
  \begin{subfigure}[t]{\eesize\linewidth}
    \includegraphics[width=1\textwidth]{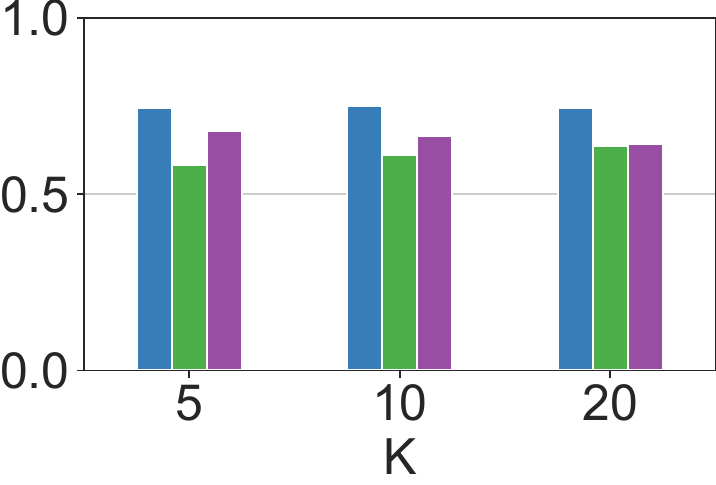}
    \caption{Movies}
    \label{fig:aps-movies}
  \end{subfigure}
  \begin{subfigure}[t]{\eesize\linewidth}
    \includegraphics[width=1\textwidth]{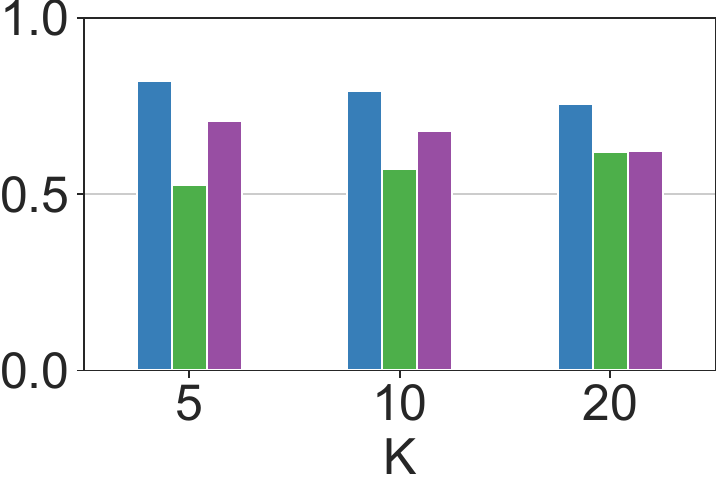}
    \caption{UCF101}
    \label{fig:aps-ucf101}
  \end{subfigure}
  \caption{
    Average pairwise similarity (APS) for \sn and the baselines.
    Lower is better.
  }
  \label{fig:aps-all}
\end{figure*}

\section{Evaluation} \label{sec:evaluation}
Our evaluation aims to answer the following research questions:
\begin{enumerate}
  \item Can \sn be used to produce relevant and semantically diverse \topk
    results? (\msec{eval-e2e})
  \item How does \sn compare to existing approaches for \topk video analytics?
    (\msec{eval-performance})
  \item How do \sn's different components contribute to producing \topk
    results? (\msec{eval-ablation})
  \item Can \sn automatically remove low-quality data? (\msec{eval-quality})
  \item How does using \clip's semantic embeddings compare to using pixel-wise
    similarity to diversify results? (\msec{eval-vissim})
\end{enumerate}

\subsection{Setup} \label{sec:eval-setup}
\minihead{System Configuration} We deployed
\sn on Google Cloud Platform (GCP).
We used a n1-highmem-16 instance (16 vCPUs, 104 GB of DRAM) with an
NVIDIA T4 GPU.
This instance features Intel Xeon E5-2699 v4 CPUs operating at 2.20GHz, Ubuntu
20.04 with 5.15.0 kernel.
We use a pretrained ViT-B/32 \clip model~\cite{clip}.

\minihead{Queries and Datasets}
Table~\ref{tab:datasets} shows the datasets, number of frames, and predicates used to
evaluate \sn. Each predicate is passed as a single prompt to \sn.
BDD~\cite{bdd} is a dashcam footage dataset used for evaluating models for
self-driving vehicles.
CCT~\cite{cct} is an animal trap dataset used for ecological applications.
Movies~\cite{moviesdataset} is a collection of $\sim$1 hour long films that can
be studied for performing actor demographic analysis.
TV news data has been used to explore a decade of US cable
news~\cite{tvarch21,tvanalyzing}.
UCF101~\cite{ucf} is a collection of activities used for action recognition.
We constructed a diverse set of queries with both single and multiple predicates
drawn from recent work in video
analytics~\cite{viva,figo,blazeit,noscope,pp,core,tasti}.

\minihead{Metrics}
We use the following metrics in our evaluation:
\begin{itemize}
  \item \textit{Retrieval Mean Average Precision (MAP)}: to measure whether \sn
    can return relevant results and rank relevant results effectively, we use
    retrieval MAP.
    It is a commonly-used metric in information retrieval for evaluating
    ranking systems~\cite{pytorch,rmap}.
    It measures the order that matching results are returned to a user.
    For a ranked set of $K$ frames returned, the average precision (AP) is
    computed by:
    \begin{small}
      \begin{equation*}
        AP = \frac{1}{RF} \sum_{k=1}^{K} P(k) r(k)
      \end{equation*}
    \end{small}
  Where $RF$ is the total number of matching frames returned, $P(k)$ is the
  precision up to $k$ frames, $r(k)$ is the relevance of the $k^{th}$ frame (1
  if the predicate is in the frame, if not 0).
  We can then compute MAP across all queries in a dataset as:
  \begin{small}
    \begin{equation*}
      MAP = \frac{1}{Q} \sum_{q=1}^{Q} AP(q)
    \end{equation*}
  \end{small}
  Where $AP(q)$ is the AP for query $q$, and $Q$ is the number of queries.
  MAP ranges from 0 to 1, where a higher MAP indicates relevant results are
  ranked higher. \sn maximizes MAP.
  \item \textit{Average Pairwise Similarity (APS)}: to measure how distinct
  each of the ranked results are from each other, we use APS.
  This metric is typically used in measuring similarity in data clustering
  problems~\cite{aps}.
  APS computes the average cosine similarity between pairs of frames returned in
  the \topk results.
  APS ranges from 0 to 1, where a lower APS indicates frames are dissimilar
  from one another.
  APS will vary more in video datasets where there is larger visual differences
  like movies and interviews (\eg many different people and scenes).
  It will have lower variation for datasets with a fixed view (\eg camera trap)
  or many similar frames (\eg dashcam footage). \sn minimizes APS.
\end{itemize}

\minihead{Baselines}
We evaluate \sn against the following baselines.
For the VLM-based baselines, neither baseline uses additional label sets,
synonyms, or low-quality terms as prompts, consistent with how we analyzed
\clip in Section~\ref{sec:motivation}:
\begin{itemize}
  \item \blrel: this baseline ranks frames based on the cosine similarity of
    the input query embedding and frame embeddings using \clip out-of-the-box.
    It represents what a user would expect if all frame embeddings were stored in
    a vector database and the K-nearest frames to the query prompt were retrieved.
    This baseline maximizes MAP but does nothing to minimize APS.
  \item \bldiv: this baseline ranks frames based on the pairwise cosine
    similarity of frame embeddings produced by \clip out-of-the-box.
    It represents what a user would expect if they wanted to retrieve results that
    were as semantically different as possible from each other.
    This baseline minimizes APS but does nothing to maximize MAP.
  \item VIVA: this baseline represents state-of-the-art video analytics systems
    like VIVA, BlazeIt, and Everest. It executes queries with a combination of
    fast, low accuracy proxy models and slow, more accurate larger models and
    uses pixel difference for similarity analysis. This baseline represents what
    a user would expect using existing non-VLM based systems.
\end{itemize}

\begin{figure}[t]
  \centering
  \begin{minipage}[t]{1\columnwidth}
    \centering
    \includegraphics[width=1\columnwidth]{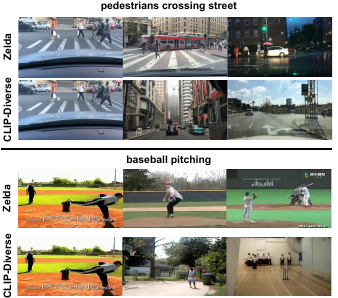}
    \caption{
      Top-K results from \sn compared to \bldiv for two queries.
      \sn produces relevant and more semantically diverse \topk results
      compared to VLMs out-of-the-box.
    }
    \label{fig:zelda-qualitative}
  \end{minipage}
\end{figure}

\subsection{Producing Diverse, Relevant Ranked Results} \label{sec:eval-e2e}
We first show \sn is able to produce relevant, diverse \topk results.
We compare \sn to \blrel and \bldiv.
\blrel optimizes for higher MAP while \bldiv optimizes for lower APS.
Similar to prior work, we consider \topk values of 5, 10, and 20~\cite{value,fixy}.

Figures~\ref{fig:map-all} and~\ref{fig:aps-all} show the MAP and APS for each
of the five datasets, respectively.
We generally see \blrel has the highest MAP, but also the highest APS.
Similarly, \bldiv generally has the lowest MAP, but also the lowest APS.
\sn provides the best of both baselines: on average equivalent to 
\texttt{\clip}-\texttt{Relevant} %
in MAP, but has better diversity (\apsoverbrel higher APS on average).
Compared to \bldiv, \sn's APS is \apsoverbdiv higher on average but achieves
higher MAP by 1.15$\times$ on average.
Figures~\ref{fig:zelda-qualitative} shows results for \sn and \bldiv for 2 queries.
We see \sn produces more relevant but still semantically diverse results.
For \q{baseball pitching}, \bldiv favors diversity which returns very few
relevant results. Conversely, \sn finds the right balance automatically.
We now discuss how \sn compares to the two baselines for each dataset in
Figures~\ref{fig:map-all} and~\ref{fig:aps-all}.

\miniheadit{BDD}
\sn's improved MAP over \bldiv (1.11$\times$ across all K) primarily comes from
the query for finding pedestrians crossing street, where non-matching frames
tend to be empty crosswalks or intersections.
\blrel's reduced accuracy comes from showing these non-matching frames early,
and since many look similar it ranks them higher.
Despite the majority of frames containing cars and vehicles, \sn still shows a
diverse set of results: only 1.02$\times$ worse APS across all K compared to
\bldiv and 1.15$\times$ better than \blrel.

\miniheadit{CCT}
Raccoon and deer are rare animals in CCT and can be difficult to identify
especially at night.
They can also be mistaken as other animals (\eg coyotes or cats).
Thus, as K increases, the MAP decreases for \sn and the baselines.
For top-5 results, \bldiv's MAP is the highest since it only shows one deer as
the top result.
\sn and \blrel show additional deer results, but not all consecutively as the
top ranked ones, which slightly decreases their MAP$@$5.
Across all K, \sn's MAP is within 1.01$\times$ of equivalent to
\bldiv on average. \sn's APS is 1.11$\times$ better than \blrel across all K.

\miniheadit{News}
As shown in Figure~\ref{fig:ootb-qualitative}, VLMs return relevant results for
queries like \q{Jake Tapper}.
However, \blrel produces results that are visually similar.
\sn and
\texttt{\clip}-\texttt{Diverse} %
produce a more diverse set of results by pruning semantically
similar frames.
For example, both \sn and \bldiv show results with polls featuring Bernie
Sanders.
This improves \sn's APS by 1.28$\times$ across all K compared to \blrel.
By ranking more confident results higher after removing similar frames, \sn
improves MAP over \bldiv by 1.12$\times$ across all K.

\miniheadit{Movies}
Similar to BDD, \sn's improvements for MAP over both baselines primarily comes
from the query for finding pedestrians crossing street.
For the cars query, we also observe multiple blurry frames identified as cars
and ranked high.
By using \q{blurry} as an additional prompt, \sn was able to eliminate these,
thus improving MAP (explored further in Section~\ref{sec:eval-quality}).
Even after removing these low-quality frames, \sn improves APS by
1.14$\times$ across all K compared to \blrel.

\miniheadit{UCF101}
Both \sn and \bldiv produce more diverse results than \blrel (\eg other sports
for archery). This enables \sn to improve APS by 1.19$\times$ on average compared to
\texttt{\clip}-\texttt{Relevant}. %
By ranking the most relevant results at the top, \sn's MAP is
1.41$\times$ higher than \bldiv across all K.

\subsection{Performance versus Existing Systems} \label{sec:eval-performance}
We next show by using VLMs, \sn improves latency by an order of
magnitude over state-of-the-art systems and optimizations in video analytics.
We consider a baseline implemented with VIVA that is representative of
BlazeIt~\cite{blazeit} and Everest~\cite{everest}.
In this baseline, queries are executed using a combination of fast, low
accuracy proxy models and slow, more accurate models.
This baseline aims to use the fast model as much as possible.
If the proxy model's confidence is lower than a fixed threshold, it will
fall back to the more accurate, slow model.
We compare 2 scenarios:
(1) \vivaa: the threshold is set higher (0.9), which will call the accurate model
more often.
(2) \vivaf: the threshold is set lower (0.2), which will call the accurate model less
often and instead rely primarily on the fast model's results.
We consider a single predicate query --- \q{car} --- and a multiple predicate
query --- \q{cars at night}.
For \q{car}, the baselines use the same YOLO-fast and YOLO-accurate models
from Table~\ref{tab:motivation-throughput} to represent a fast, less accurate
model and a slow, more accurate model.
For \q{night}, the baselines use an SVM time of day detector previously used by
VIVA~\cite{hints}.

\begin{figure}[t]
  \centering
  \begin{subfigure}[t]{1.0\linewidth}
    \includegraphics[width=1\textwidth]{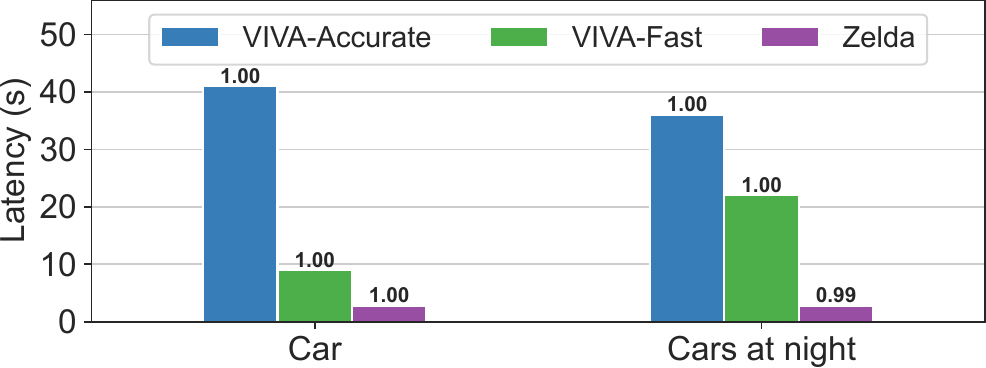}
  \end{subfigure}
  \caption{
      \sn compared to \vivaa and \vivaf.
      AP@20 is annotated above each bar showing almost identical retrieval
      accuracies.
      \sn improves performance by up to 10.4$\times$ compared to
      state-of-the-art video analytics systems.
  }
  \label{fig:baseline-perf}
\end{figure}

Figure~\ref{fig:baseline-perf} shows the latencies of the baselines compared to
\sn.
Annotated above each bar is the AP$@$20 where we see all baselines have almost
identical accuracies.
As expected, \vivaa is the slowest because the confidence requirement is high
and most frames are labelled by the more accurate model.
\vivaf is faster than \vivaa, but is slower than \sn.
\sn benefits from using \clip: a single, fast VLM.
On average \sn is \perfaccurate faster than \vivaa and \perffast faster than
\vivaf.
We also post-process \vivaa and \vivaf's \topk results with \clip to compute
their APS$@$20.
Both baselines have equivalent APS$@$20 per-query.
Since neither baseline has a way to diversify results, \sn's APS$@$20 is
1.11$\times$ better for the car query and 1.02$\times$ better for the cars at
night query.

\subsection{Ablating \sn} \label{sec:eval-ablation}
We now show how each of \sn's different components contribute to producing
relevant and semantically diverse \topk results.
We show three different ablated versions of \sn:
(1) \blrel: ranked based on cosine similarity of prompt,
(2) \texttt{+Prompting}: prompting with LM-generated synonyms, the LVIS label
set, and data quality descriptors (\dcleaners),
and (3) \texttt{+Diversity Rank}: ranking results after removing frames that
are visually similar.
We focus on top-20 results for all five datasets.

\begin{figure}[t]
  \centering
  \begin{subfigure}[t]{1.0\linewidth}
    \includegraphics[width=1\textwidth]{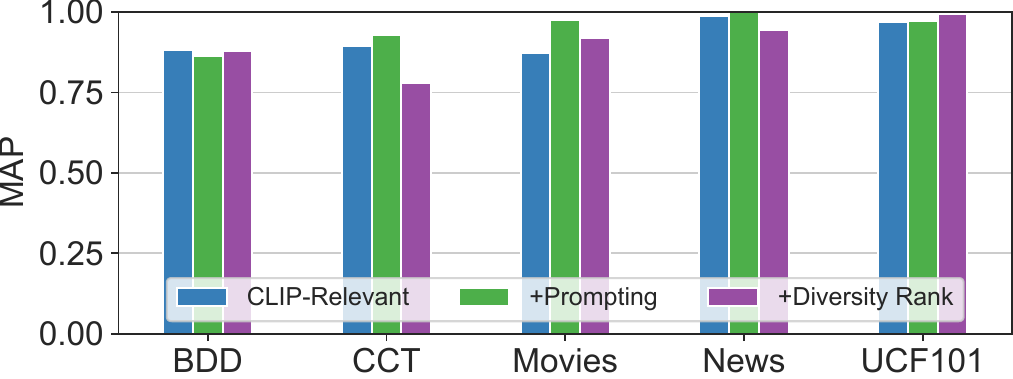}
    \label{fig:abl-map}
  \end{subfigure}
  \begin{subfigure}[t]{1.0\linewidth}
    \includegraphics[width=1\textwidth]{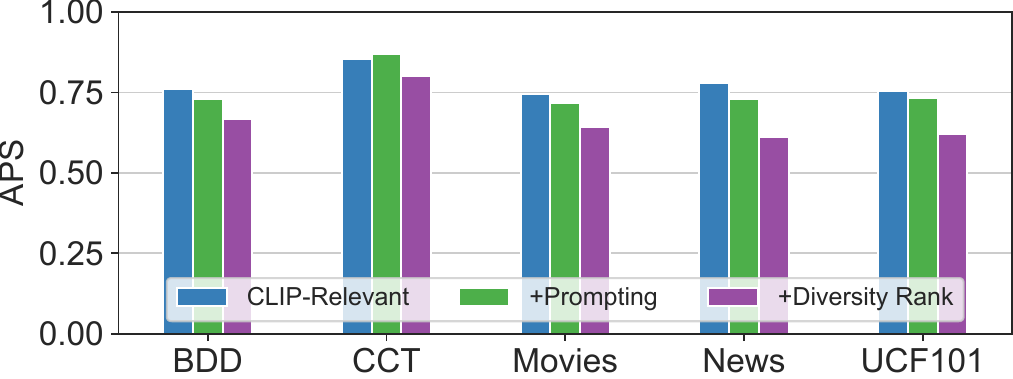}
    \label{fig:abl-aps}
  \end{subfigure}
  \caption{
    Ablated mean average precision (MAP) and average pairwise similarity (APS)
    for top-20 results across all datasets.
    Each of \sn's components help it produce diverse, relevant results with
    high MAP and low APS.
  }
  \label{fig:ablation-all}
\end{figure}

Figure~\ref{fig:ablation-all} shows the MAP and APS for the ablated results for
each of the five datasets, respectively.
Adding the LM-generated synonyms, the LVIS label set, and data quality prompts
(\texttt{+Prompting}) improves MAP by 1.04$\times$ on average.
This also slightly improves APS by 1.03$\times$.
As described in \msec{sys-cand-gen}, \sn uses label sets and softmaxes the output
to \q{discriminate} between nearby objects in the embedding space.
For example, animals can typically be confused with each other in the CCT dataset.
Two nearby animals to raccoon in the LVIS label set are cat and squirrel, which
serve as discriminators.
This improves the AP$@$20 for raccoon from 0.84 to 0.94.
By ranking results after removing frames that are similar to each other
(\texttt{+Diversity Rank}), \sn's MAP slightly decreases compared to
\texttt{+Prompting}.
However, \sn's APS is 1.15$\times$ better than \texttt{+Prompting}.

Table~\ref{tab:synonym} shows the average precision for top-10
results in the News dataset for 4 different queries.
Adding LM-generated synonyms (\texttt{+Prompting}) to the set
of prompts improves precision by 1.18$\times$ on average.
Some of these queries are general categories (\eg \q{water sport})
and the language model generates prompts that are more
specific and descriptive (\eg \q{surfing}) which
improves precision by up to 1.25$\times$.

\newcommand{\ablatefigdir}{tex/figures/ablation}

\begin{table}[t]
    \caption{
        Ablated average precision in top-10
        results for queries where \sn
        generates synonym prompts.
    }
    \resizebox{0.65\linewidth}{!}{%
        \begin{tabular}{l|cc}
            \toprule
            \multirow{2}{*}{\textbf{Query}} & \multicolumn{2}{c}{\textbf{Average Precision@10}}                                      \\
                                            & \textbf{\blrel}                                   & \textbf{\texttt{+Synonym Prompts}} \\ \hline
            workplace                       & \phantom{0}0.8                                    & \bf{1.0}                           \\
            zoo animal                      & \phantom{0}0.24                                   & \bf{0.33}                          \\
            water sport                     & \phantom{0}0.79                                   & \bf{1.0}                           \\
            food                            & \phantom{0}\bf{0.82}                              & 0.79                               \\
            \bottomrule
        \end{tabular}
    }
    \label{tab:synonym}
\end{table}

\subsection{Removing Low-Quality Data} \label{sec:eval-quality}
By prompting \sn with terms like \dcleaners, we show \sn can remove low-quality
results.
This improves ease-of-exploration for users, while maintaining high MAP.
We compare \sn to \bldiv, \blrel, and \texttt{No dataquality}: a version of
\sn that does not include the low-quality terms.
We consider queries in the BDD, Movies, and UCF101 datasets where there were at
least two blurry frames from \texttt{No dataquality}, and again focus on
top-20 results.

\begin{figure}[h]
  \centering
  \begin{minipage}[t]{0.95\columnwidth}
    \centering
    \includegraphics[width=1\columnwidth]{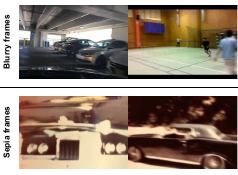}
    \caption{
      \sn is able to identify frames prompted with \q{blurry, sepia} that are
      consequently pruned.
    }
    \label{fig:zelda-quality}
  \end{minipage}
\end{figure}

Table~\ref{tab:dataclean} shows the number of blurry results and the average
precision$@$20 (AP$@$20).
Figure~\ref{fig:zelda-quality} shows some example frames that were removed by
using the data quality prompts.
By prompting \sn with \dcleaners, \sn identifies and reduces the number of
low-quality results.
The term \q{blurry} is the most prevalent and has the highest cosine similarity
with our datasets.
This is expected videos can have rapid camera movements.
For queries in the BDD dataset, \sn only includes one low-quality top-20
result.
AP$@$20 is either equivalent or higher than \texttt{No dataquality} and \bldiv.
Compared to \blrel, \sn only has lower AP$@$20 for \q{cars at night at traffic
intersections} since a high-ranked blurry frame is replaced with one that
contains cars at night but not in a traffic intersection.
For queries in the Movies and UCF101 dataset, \sn reduces the number of
low-quality results.
Querying the Movies dataset for \q{car} with \blrel had a majority of low-quality
car results which also impacted its AP$@$20.
For the \q{car} query in the Movies dataset, as noted in Section~\ref{sec:eval-e2e}, \sn
improves AP by removing high-ranked blurry results (up to 1.28$\times$).
For queries in the UCF101 dataset, AP$@$20 is equivalent to \texttt{No
dataquality} because blurry frames were low-ranked results of other sports.

Figure~\ref{fig:zelda-quality} shows examples of sepia-toned frames \sn can
automatically remove by using the prompt \q{sepia}.
The left frame appears in the top-20 results of both \blrel and
\texttt{\clip}-\texttt{Diverse} %
when searching for cars in the Movies dataset, while the right frame appears in
the results of \blrel.

We loosely translated other video analytics predicates to prompts that could act
as filters and saw mixed results. Estimated object counts (\eg \q{many, a
lot, few, some}) and relative locations in the image (\eg \q{right, left, east,
west}) did poorly while colors (\eg \q{red, blue}) produced results that were
occasionally attributed to the predicate.
For example, when prompting for \q{blue cars}, the VLM would return a mix of
blue cars and cars with blue in other parts of the frame).
As the accuracy of these models increase, we expect the accuracy of prompting
VLMs with more specific terms to increase.
This will make VLMs even more useful for video analytics.

\begin{table}[h]
  \caption{
    Number of blurry frames and average precision in top-20 results.
    \texttt{No dataquality} is a version of \sn that does not include 
    prompts to remove low-quality (blurry) frames.
    \sn automatically removes low-quality results.
  }
  \resizebox{\linewidth}{!}{%
  \begin{tabular}{ll|cccc}
    \toprule
    \multirow{2}{*}{\textbf{Query}} & \multirow{2}{*}{\textbf{Dataset}} & \multicolumn{4}{c}{\textbf{\# blurry frames (Average Precision@20)}} \\
    & & \textbf{\blrel} & \textbf{\bldiv} & \textbf{\texttt{No dataquality}} & \textbf{\sn} \\ \hline
    cars              & BDD    & \phantom{0}1 (1.00) & 1 (0.89) & 3 (1.00) & 0 (1.00) \\
    vehicles          & BDD    & \phantom{0}1 (1.00) & 1 (1.00) & 2 (1.00) & 0 (1.00) \\
    \tworowslt{cars at night at}{traffic intersections} & BDD & \phantom{0}3 (0.79) & 3 (0.67) & 3 (0.63) & 1 (0.69) \\
    cars              & Movies & 16 (0.75)           & 3 (0.86) & 3 (0.92) & 2 (0.96) \\
    archery           & UCF101 & \phantom{0}0 (1.00) & 2 (0.45) & 3 (1.00) & 1 (1.00) \\
    baseball pitching & UCF101 & \phantom{0}1 (0.95) & 2 (0.43) & 2 (1.00) & 1 (1.00) \\
    \bottomrule
  \end{tabular}
}
  \label{tab:dataclean}
\end{table}

\subsection{Comparing Visual and Semantic Similarity} \label{sec:eval-vissim}
Finally, we show by leveraging the semantic understanding of VLM embeddings,
\sn diversifies \topk results and eliminates the need to perform extensive
tuning of visual difference detectors (VDDs).
As detailed in Section~\ref{sec:background}, VLMs are jointly trained with text
and images.
Their embeddings encode a rich, semantic relationship used for comparing
text and images.
We consider the case where \sn's use of cosine distance of VLM embeddings in the
result diversity algorithm (Algorithm~\ref{alg:diversity}) is replaced with a
VDD.
As is done in NoScope, the VDD computes the mean squared error between frames:
higher mean squared error (MSE) indicates frames are more dissimilar from each
other.
For this experiment, we empirically set the minimum MSE between frames to be
1.5: higher values lead to an insufficient number of top-20 frames while lower
values lead to redundant frames being selected in the top-20.
We consider all queries and datasets from Table~\ref{tab:datasets}.

\begin{figure}[t]
  \centering
  \begin{subfigure}[t]{1.0\linewidth}
    \includegraphics[width=1\textwidth]{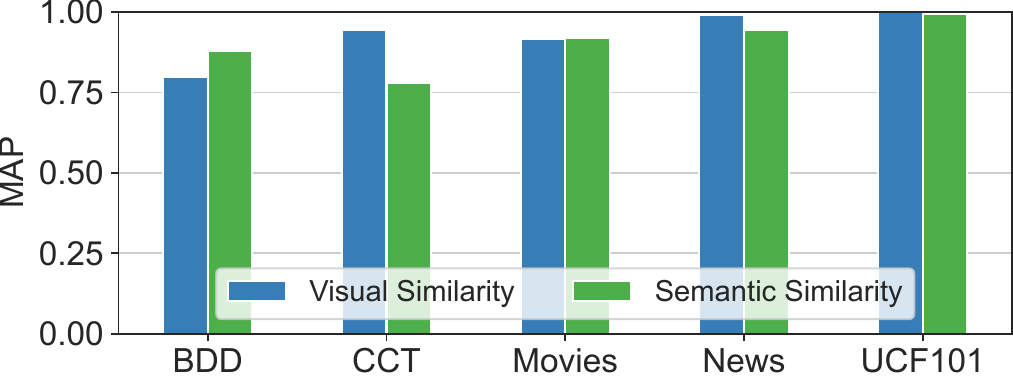}
    \label{fig:diffdet-map}
  \end{subfigure}
  \begin{subfigure}[t]{0.99\linewidth}
    \includegraphics[width=1\textwidth]{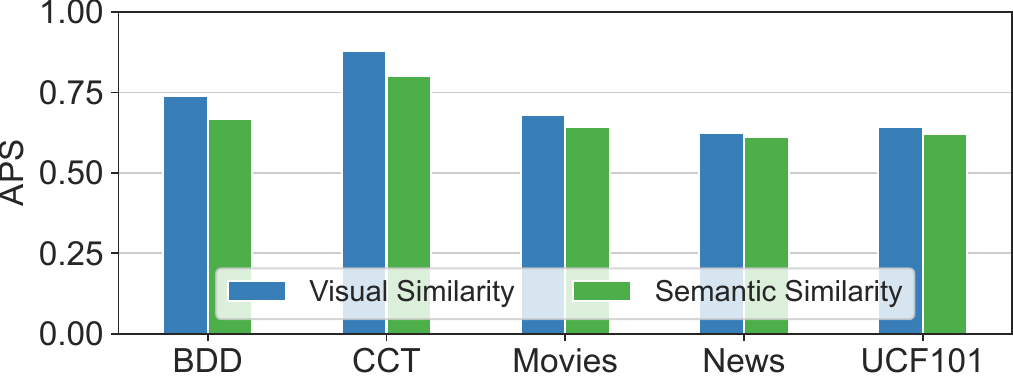}
    \label{fig:diffdet-aps}
  \end{subfigure}
  \caption{
    MAP and APS for top-20 results across all datasets when comparing visual and semantic similarity.
    Semantic similarity (used by \sn) requires less tuning than visual
    similarity and improves APS by 1.06$\times$ while maintaining high MAP.
  }
  \label{fig:diffdet-all}
\end{figure}

Figure~\ref{fig:diffdet-all} shows the MAP (higher is better) and APS (lower is
better) for top-20 queries for diversifying results using semantic versus visual
similarity.
By using semantic similarity to improve diversity, \sn improves APS by
1.06$\times$ on average.
Given the candidate results, we did not expect a significant difference because
pixel-wise similarity already distinguishes between visually distinct frames
well.
However, this shows that VLMs can further diversify the results using both
visual and semantic information about the frames without requiring an additional
model or technique because the embeddings are already available.
The MAP for the two is close (1.02$\times$ difference) since \sn ranks the
remaining frames in its final stage for both cases.
The largest improvement comes for the BDD dataset's queries --- particularly
\q{cars at night at traffic intersections} and \q{pedestrians crossing street}.
For example, with visual similarity the former query's results tends to show
diverse sets of cars driving down the street, but many of them are not at a
traffic intersection.
In contrast, using semantic similarity produces more results with stoplights and
traffic intersections.
Overall for BDD, using semantic similarity results in 1.10$\times$ and
1.16$\times$ improvement for MAP and APS respectively.
In practice, as shown by NoScope, the best MSE value can vary widely per query
and dataset.

\newcommand{\diffdetectfigdir}{tex/figures/diff_detector}

\section{Related Work} \label{sec:rel_work}
\minihead{Exploratory Video Analytics}
VOCALExplore~\cite{vocalexplore} SeeSaw~\cite{seesaw} are the most
relevant works to this paper. They also explore harnessing the capabilities of
VLMs in the context of video analytics for interactively searching a large scale
video dataset. They propose systems where a human-in-the-loop provides
feedback and annotations. VOCALExplore uses this feedback to remove low-quality
data and train specialized query-specific models. In this work, we demonstrate
VLMs produce accurate results without the need to train a query-specific model.
\sn automatically removes low-quality data without the need for manual
annotations. SeeSaw refines the results based on user feedback and an adaptive
resolution optimization that analyzes different spatial locations in a frame.
However, neither SeeSaw nor VOCALExplore consider result diversification, which
is important when presenting \topk results to users.

\minihead{Top-K Queries}
Efficient and accurate \topk query processing is a widely researched topic in
the database community~\cite{topksurvey,topx,topkdb}.
There has been recent focus on \topk querying processing for ML workloads.
Everest~\cite{everest} investigates providing probabilistic guarantees for \topk
queries for video analytics. Their work uses models with fixed classes and
requires training a model per-query.
ColBERT~\cite{colbert, colbertv2} and PLAID~\cite{plaid} are recent work that
explore \topk query processing for document retrieval using large scale
language models.
To the best of our knowledge, this is the first work that explores \topk
querying for video analytics using VLMs.

\minihead{Vision-Language Models}
\clip~\cite{clip} popularized the use of VLMs, showing state-of-the-art
results on zero-shot image classification.
Other work has explored VLMs for object detection~\cite{detic, vild},
visual-question answering and captioning~\cite{flamingo}, image-text
retrieval~\cite{filip, clipadapter}, and video understanding~\cite{videoclip,promptingforvideo}.
\sn takes advantage of the unique characteristics of VLMs as general image
and text encoders and will benefit from tbe continued research in this field.

\minihead{Model Prompting}
Segment Anything~\cite{segment}, a large segmentation model released by Meta,
makes their model \q{promptable} by taking bounding boxes, natural language, or
segment points as inputs. This shows the ongoing trend towards more
expressive interfaces for ML models.
AutoPrompt~\cite{autoprompt} shows that prompting models is challenging and
error-prone, and investigates techniques to automatically learn prompts in
embedding space.
CoOp~\cite{coop} and CoCoOp~\cite{cocoop} investigate turning additional
context one might add to a prompt into learnable vectors.
They also show improvement gains over manually crafted prompts.
Analagous concept generation (ACG) shows that language models can generate
synonyms~\cite{analogy}.
The techniques developed in our work would complementarily benefit from these
techniques that improve the accuracy of prompting.
They may also enable different ways of prompting a model.

\section{Conclusion}
Today's video analytics systems limit expressivity, need multiple models to
match query predicates, require complex optimizations to achieve high accuracy
and low latency, and return redundant and low-quality results.
In this work, we show VLMs improve general expressivity, can be a general
purpose model, and can improve performance.
We showed VLMs are easy to use, widely applicable to many queries, highly
accurate, and fast for both single and multiple predicate queries.
To reduce the number of results a user needs to interpret, we built \sn: a
system for \snverbose.
Using a natural language input, \sn generates relevant candidate frames
without any user annotation or query specific training.
\sn automatically removes low-quality frames using VLMs and uses VLM-generated
semantic embeddings to improve diversity over existing approaches while still
returning relevant results.
\sn achieves higher retrieval mean average precision (up to \mapoverbdiv)
compared to out-of-the-box VLMs while improving similarity between frames by
up to \apsoverbrel.
\sn is on average~\perfavg (up to \perfaccurate) faster at retrieving results
than a state-of-the-art video analytics system.

\bibliographystyle{ACM-Reference-Format}
\bibliography{viva-vlm}

\end{document}